\documentclass{article}
\usepackage{spconf,amsmath,graphicx}
\usepackage[utf8]{inputenc}
\usepackage{lmodern,textcomp}
\usepackage{graphics}
\usepackage{multicol}
\usepackage{lipsum}
\usepackage{color}
\usepackage{amsthm}
\usepackage{amssymb}
\usepackage{mathtools}
\usepackage{amsbsy}
\usepackage{setspace}
\usepackage{float}
\usepackage{lettrine}
\usepackage{newclude}
\usepackage[normalem]{ulem}
\usepackage{latexsym}
\usepackage{multirow}
\usepackage{rotating}
\usepackage{booktabs}
\usepackage{wasysym}
\usepackage{mathrsfs}
\usepackage{bbm} 
\usepackage{epsfig,cite,amsfonts,psfrag}
\usepackage{apptools}
\usepackage{graphicx}
\usepackage[table]{xcolor}
\usepackage{chngcntr}
\usepackage{blindtext}
\usepackage{optidef}
\usepackage{accents}
\usepackage{dsfont}
\usepackage{siunitx}
\usepackage{overpic}

\title{Quantization-Aware Precoding for MU-MIMO with Limited-Capacity Fronthaul}

\name{Yasaman Khorsandmanesh, Emil Björnson, and Joakim Jaldén \thanks{This work was supported by the Knut and Alice Wallenberg Foundation. E.~Björnson is also with Department of Electrical Engineering, Linköping University, Linköping, Sweden.} }\address{School of Electrical Engineering and Computer Science, KTH Royal Institute of Technology, \\ Stockholm, Sweden.  Email: \{yasamank, emilbjo, jalden\}@kth.se}

\begin{document}
\maketitle
\ninept

%%%%%%%%%%%%%%%%%%%%%%%%%%%%%%
\begin{abstract}
Base stations in 5G and beyond use advanced antenna systems (AASs), where multiple passive antenna elements and radio units are integrated into a single box. A critical bottleneck of such a system is the digital fronthaul between the AAS and baseband unit (BBU), which has limited capacity. In this paper, we study an AAS used for precoded downlink transmission over a multi-user multiple-input multiple-output (MU-MIMO) channel.
First, we present the baseline 
quantization-unaware precoding scheme created when a precoder is computed at the BBU and then quantized to be sent over the fronthaul.
We propose a new precoding design that is aware of the fronthaul quantization. We formulate an optimization problem to minimize the mean squared error at the receiver side. We rewrite the problem to utilize mixed-integer programming to solve it. The numerical results manifest that our proposed precoding greatly outperforms quantization-unaware precoding in terms of sum rate. 
\end{abstract}


%%%%%%%%%%%%%%%%%%%%%%%%%%%%%%%%%%%%%%%%%%
\section{Introduction} \label{sec:intro}

\begin{figure}[t!]
  \centering
   \begin{overpic}[scale=0.36]{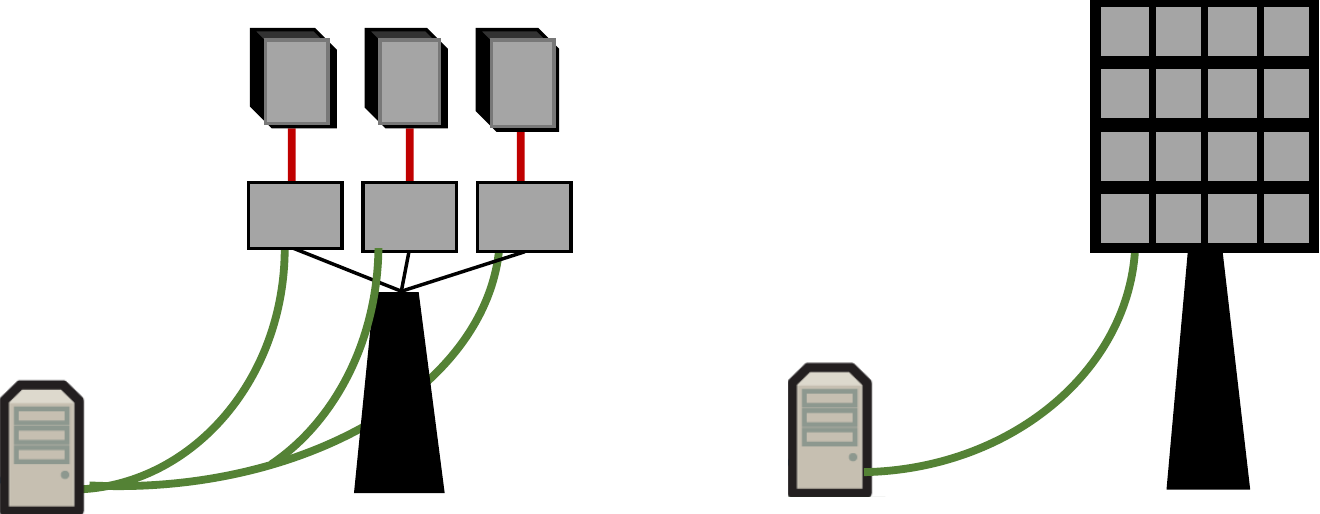}
  \put(-3,11){BBU}%
  \put(57,12){BBU}%
  \put(18.4,21){\scriptsize RU}%
  \put(35.5,21){\scriptsize RU}%
  \put(27,21){\scriptsize RU}%
  \put(14,40){Passive AEs}%
  \put(85,40){AAS}%
  \put(-13,-7){(a) The traditional BS}%
  \put(59,-7){(b) The 5G BS}%
   \end{overpic} \vspace{1mm}
\caption{The base stations' configuration (a) shows the analog connection and (b) digital fronthaul.}
\label{fig:systemmodel} \vspace{-4mm}
\end{figure}
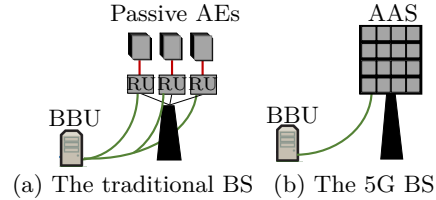

Multi-user multiple-input multiple-output (MU-MIMO) is a classical technology for spatial multiplexing of user equipments (UEs) on the same time-frequency resource by utilizing a base station (BS) with multiple antennas \cite{Swales1990a,Gesbert2007a}.
MU-MIMO has been supported by several standards, but 5G networks are the first to make widespread use of it \cite{Parkvall2017a}. A key reason for the slow adoption is that a traditional BS contains one baseband unit (BBU) and then two boxes per antenna: one passive antenna element (AE) and one radio unit (RU), as depicted in Fig.~\ref{fig:systemmodel}(a). However, 5G BSs integrate all AEs and RUs into a single enclosure, called an \emph{advanced antenna system (AAS)} \cite[Chapter 1]{asplund2020advanced} shown in Fig.~\ref{fig:systemmodel}(b).
This hardware evolution has made massive MU-MIMO practically feasible \cite{Bjornson2019d} and enables the BBU to be virtualized in the cloud. A new implementation bottleneck is the digital fronthaul between the AAS and BBU, which needs a capacity proportional to the number of antennas. This interface must carry received uplink signals (to be decoded at the BBU) and precoded downlink signals, which are computed at the BBU. In this paper, we propose a new linear block-level precoding technique that is aware of the necessary fronthaul quantization.

%%%%%%%%%%%%%%%%%%%%%%%%%%%%%%%%%%%%%%%%%%%%%
\subsection{Related Works}

The impact of impairments in analog hardware on MU-MIMO systems has received much attention in prior literature (see e.g., \cite{wenk2010mimo,Zhang2012a,Bjornson2014a,Mollen2018a,aghdam2020distortion}). There are related works on quantization distortion caused by low-resolution analog-to-digital converters in the uplink \cite{mollen2016uplink,studer2016quantized} and 
low-resolution digital-to-analog converters (DAC) in the downlink \cite{jacobsson2017quantized,jacobsson2019linear,castaneda2019finite}. A key characteristic of these prior works is that the distortion is created in the RU, the analog domain, or the converters. This implies that the transmit signal obtained after precoding is distorted. One way to handle low-resolution DACs is to perform symbol-level precoding \cite{jacobsson2016nonlinear,tsinos2018symbol}, where a new precoder is selected for each symbol vector to minimize signal distortion. This approach requires much more fronthaul signaling and computational complexity than symbol-level precoding. Another related line of work is quantized feedback \cite{Love2008a,jindal2006mimo}, where UEs estimate and feed back their channels to the BS. The precoding is computed based on the quantized channels, which leads to extra unremovable interference even if zero-forcing is used \cite{jindal2006mimo}, but the key difference is that the quantization appears before the precoding.
The effect of limited fronthaul capacity is studied in \cite{simeone2009downlink,park2014fronthaul, parida2018downlink} among others, but the focus was not on precoding design. To our knowledge, this is the first paper to analyze the precoding distortion that occurs over the digital interface between the BBU and AAS, where the key difference is that the precoding matrix is quantized before the transmit signal is computed; that is, before the quantized precoding matrix is multiplied with the data symbols at the AAS. An extended version of this paper is available in \cite{Yasaman2022}.

%%%%%%%%%%%%%%%%%%%%%%%%%%%%%%%%%%%%%%%%%%%%%
\subsection{Contributions}

In this paper, we consider MU-MIMO precoding quantization over a limited-capacity fronthaul connection. Since the data symbols originate from a finite-resolution codebook and change much more rapidly than the precoding matrix, an efficient implementation will send these quantities separately over the fronthaul so that only the precoder is quantized. We formulate and solve a novel quantization-aware precoding problem, where the communication performance after quantization is maximized.
The main contributions are:

\begin{itemize}
    \item Inspired by practical AAS implementation, we formulate a new downlink precoding framework where the precoding matrix is quantized when sent over the limited-capacity fronthaul from the BBU to the AAS, while data symbols are inherently quantized.
    
    \item We formulate a quantization-aware precoding problem, where the precoder is selected to minimize the mean-squared error (MSE) at the receiver side. We use mixed-integer programming (MIP) \cite{wolsey2007mixed,gurobi2020reference} to solve it to global optimality for a fixed precoding coefficient. % that minimizing the  MSE.
    
    \item We provide numerical results to show the benefits of quantization-aware precoding over the quantization-unaware baseline. We describe how the number of quantization levels and UEs affect performance.
    
\end{itemize}

%%%%%%%%%%%%%%%%%%%%%%%%%%%%%%%%%%%%%%%%%%%%%%%%%%%%%%%%%%%%%%%%
\section{System Model} \label{sec2}

We consider the downlink data transmission in a single-cell MU-MIMO system, where a BS equipped with an AAS with $M$ antenna-integrated radios serves $K$ single-antenna UEs on the same time-frequency resource. The AAS is connected to a BBU through a digital fronthaul link with limited capacity, as depicted in Fig.~\ref{fig:systemmodel}(b). Hence, any signal that is sent over the fronthaul must be quantized to finite resolution. Each transmitted signal vector is the product between a precoder matrix and a vector with data symbols, where the former is assumed fixed for the duration of the transmission while the latter changes at the symbol rate.
The BBU computes a precoder based on channel state information (CSI) and then forwards it to the AAS. As the data symbols represent bit sequences from a codebook, we can send them without quantization errors and map them to modulation symbols at the AAS. However, the precoder matrix normally contains arbitrary complex-valued entries and must be quantized before being sent over the limited-capacity fronthaul. The quantized matrix is then multiplied with the UEs' data symbols at the AAS, and finally the product is transmitted wirelessly.

Before analyzing the proposed quantization-aware  precoder in Section \ref{sec3solve}, in the following subsections, we first introduce the considered channel model. Then, conventional uniform quantizer-mapping and the quantization operator are described.

%%%%%%%%%%%%%%%%%%%%%%%%%%%%%%%%%%%%%%%%%%%%%%%%%%%%%%
\subsection{Channel Model}\label{sec2chamodel}

As the main focus of this work is on quantization effects, we assume the BBU has perfect CSI and neglect all other potential hardware impairments. The imperfect CSI case will be considered in future work. The downlink system model can be written as
\begin{equation}
\textbf{y}=\textbf{H}\textbf{P}\textbf{s}+ \textbf{n},
\end{equation}
where $\textbf{y} = [y_1, \ldots, y_K]^\text{T} \in \mathbb{C}^{K} $ contains the received signals at all UEs and $y_k \in \mathbb{C}$ denotes the signal received at the $k$-th UE. The downlink channel matrix $\textbf{H} \in \mathbb{C}^{K \times M}$ has entries  $h_{k,m}$ for $k=1, \ldots ,K$ and $m=1, \ldots ,M$. It represents a narrowband channel and might be one subcarrier of a multi-carrier system. The vector $\textbf{n} \in \mathbb{C}^{K}$ represents i.i.d. additive white complex Gaussian noise with zero mean and variance $N_0$. The vector $\textbf{s} = [s_1, \ldots, s_K]^\text{T}\in \mathcal{O}^{K}$ contains the information, where $s_k $ denotes the random data symbol intended for UE $k$ and is normalized to unit power. Here, $\mathcal{O}$ is the finite set of constellation points (e.g., a conventional QAM alphabet). The quantized precoder matrix is denoted by $\textbf{P} \in \mathcal{P}^{M \times K}$, where the set of quantization alphabet $\mathcal{P}$ coincides with the complex numbers set $\mathbb{C}$ in the case of infinite resolution. We denote $\mathcal{L}= \{ l_0, \ldots ,l_{L-1} \}$ as the set of real-valued quantization labels. We assume that the same quantization alphabet is used for the real and imaginary parts. Under these assumptions, the entries of the precoded vector $\textbf{P}$ are $p_{m,k} = l_{R} + jl_{I}$ where $l_{R},l_{I} \in \mathcal{L}$. The number of quantization levels is denoted by $L=|\mathcal{L}|$, $N=\log_2 (L)$ refers to the number of quantization bits per real dimension and the set of complex-valued precoder outputs for each antenna is $\mathcal{P}=\mathcal{L} \times \mathcal{L}$.

The precoder matrix $\textbf{P}$ must satisfy the power constraint 
\begin{equation}
    \left \Vert \textbf{P} \right \|_{\mathrm{F}}^2 \le  q, \label{eq:powercons}
\end{equation}
where $\left \Vert \textbf{P} \right \|_{\mathrm{F}}$ denotes the Frobenius norm and $q$ is the maximum average transmit power of the downlink signals. To estimate the transmitted information symbol $\hat{s}_k \in \mathbb{C}$, we assume that the $k$-th UE scales the received signal $y_k$ by the precoding factor $\beta\in \mathbb{C}$ as $\hat{s}_k = \beta y_k$ with the goal of minimizing the MSE $\mathbb{E}[  \left \Vert \textbf{s} -  \hat{\textbf{s}} \right \|^2]$, where $\hat{\textbf{s}} = [\hat{s}_1, \ldots, \hat{s}_K]^\text{T}$. Specifically, by using the same scalar for all UEs, we will equalize their performance \cite{jacobsson2016nonlinear}.

\vspace{-1mm}
%%%%%%%%%%%%%%%%%%%%%%%%%%%%%%%%%%%%%%%%%%%%%%%%%%%%%%%%%%%
\subsection{Quantized Precoding} \label{sec2quantize}

When it comes to quantized precoding, we will consider two cases. The naive baseline approach is to first design a precoder $\textbf{P}_\textrm{ideal}$ based on CSI and then quantize it, which we call \emph{quantization-unaware precoding}.
We notice that there are prior works that consider a similar procedure \cite{jacobsson2016nonlinear,jacobsson2017quantized}, but quantize the product $\textbf{P}_\textrm{ideal}\textbf{s}$ and not just the precoder $\textbf{P}_\textrm{ideal}$. An alternative is to find the quantized precoder that minimizes the MSE, based on both CSI and knowledge of the quantization alphabet, which we call \emph{quantization-aware precoding}. We will now present the quantization-unaware precoding approach and then the proposed quantization-aware precoding scheme will be described in Section \ref{sec3solve}.

In quantization-unaware precoding, the precoder matrix $\textbf{P}$ is designed without taking the limited-capacity fronthaul effect into account. The precoder is designed based on CSI as if there would be no quantization and then a uniform quantization scheme is applied. The precoder matrix $\textbf{P} = \alpha \hat{\textbf{P}} = \alpha \mathcal{Q}(\textbf{W})$ describes the limited-capacity fronthaul effect, where $\textbf{W}$ is the preferred precoder matrix without quantization  and $\mathcal{Q}(\cdot) : \mathbb{C}^{M \times K} \to \mathcal{P}^{M \times K}$
denotes the quantizer-mapping function. Since uniform quantization is normally used in practice, we model $\mathcal{Q}(\cdot)$ as a symmetric uniform quantizer with step size $\Delta$. 
In general, the average power in the quantized signal is not preserved. 
As the condition $\left \Vert \textbf{W} \right \|_{\mathrm{F}}^2 =  q$ does not imply $\left \Vert \textbf{P} \right \|_{\mathrm{F}}^2 =  q$ and for assurance that the power constraint (\ref{eq:powercons}) is satisfied with equality, the output of the quantizer is scaled by a constant $\alpha = \sqrt{q/ \Vert \hat{\textbf{P}} \|_{\mathrm{F}}^2}$ at the AAS.

Each entry of the quantization labels $\mathcal{L}$ is defined as 
\begin{equation}
    l_z  = \Delta  \Big( z- \frac{L-1}{2}  \Big), \quad z=0, \ldots, L-1.
\end{equation}
Furthermore, we let $\mathcal{T} = \{ \tau_0, \ldots, \tau_L \}$, where $-\infty = \tau_0 < \tau_1 < \ldots < \tau_{(L-1)} < \tau_{L} =\infty$, specify the set of $L + 1$ quantization thresholds. For uniform quantizers, the quantization thresholds
are 
\begin{equation}
    \tau_z =  \Delta  \Big( z- \frac{L}{2}  \Big), \quad z=1, \ldots, L.
\end{equation}
The quantizer function $\mathcal{Q}(\cdot)$ can be uniquely described by the set of quantization labels $\mathcal{L} = \{ l_z | z=0, \ldots, L-1 \}$ and the set of quantization thresholds $\mathcal{T}$. The quantizer maps $w_{m,k} \in \mathbb{C}$ into the quantized output $p_{m,k}= l_r + jl_i$ if
$\mathfrak{R}\{ w_{m,k} \} \in  [\tau_r,\tau_{r+1})$ and $\mathfrak{I}\{ w_{m,k} \} \in  [\tau_i,\tau_{i+1})$.
The step size $\Delta$ of the quantizers should be chosen to minimize the distortion between the quantized and unquantized vector. The optimal step size $\Delta$ depends on the minimum MSE distribution of the input, which in our case depends on the precoding scheme. Since the distribution of the precoder elements is generally unknown, we set the step size to minimize the distortion under the maximum-entropy assumption that the per-antenna input to the quantizers is $\mathcal{CN}(0,q/M)$ distributed. The corresponding optimal step size was found in \cite{hui2001unifquantized}.

%%%%%%%%%%%%%%%%%%%%%%%%%%%%%%%%%%%%%%%%%%
\section{Quantization-aware Precoding}\label{sec3solve}

The quantization-unaware precoding scheme is clearly not the optimum quantized precoder since it neglects the quantization effect, which leads to extra interference and reduced beamforming gain. 
For example, canceling all interference using zero-forcing is optimal when the signal-to-noise ratio (SNR) is high and there is no quantization \cite{joham2005linear}, while it will not be the case in our setup.
In this section, we propose a scheme that finds an optimum quantization-aware precoder that minimizes the MSE between the received signal and the transmitted symbol vector  $\textbf{s}$ under the power constraint \eqref{eq:powercons}.
%%%%%%%%%%%%%%%%%%%%%%%%%%%%%%%%%%%%%%%%%%%
\subsection{Optimal Quantized Precoding}
We formulate the precoder optimization problem as

\begin{mini}|l|
	  {\textbf{P}\in \mathcal{P}^{M \times K} , \beta \in \mathbb{C} }{\mathbb{E}[  \left \Vert \textbf{s} -  \beta \textbf{y}\right \|^2]}{}{}
	  \addConstraint{\left \Vert \textbf{P} \right \|_{\mathrm{F}}^2\le q.} \label{eq:mmse}
\end{mini}

For a given quantized precoder matrix $\textbf{P}$, we can compute the optimal value of $\beta$ by taking the Wirtinger derivative and equate it to zero, which yields 
\begin{equation}
    \beta^{\mathrm{Opt}} = \frac{\mathrm{tr} (\textbf{P}^H\textbf{H}^H)}{\mathrm{tr} (\textbf{P}^H\textbf{H}^H\textbf{H}\textbf{P}) + KN_0}.\label{eq:betaopt}
\end{equation}
If we substitute (\ref{eq:betaopt}) into (\ref{eq:mmse}), our problem will be highly complex. An iterative method where we switch between optimizing $\textbf{P}$ and $\beta$ %that we first consider fixed $\beta$ and  equal to (\ref{eq:betawf}), 
is possible, but we have noticed that $\beta$ will not change much between iterations. Hence, we propose to pick a “good” $\beta$ and then solve (\ref{eq:mmse}) once for that $\beta$. 
In particular, we will choose $\beta = \beta^{\mathrm{WF}}$, where WF refers to Wiener filter. The WF precoder minimizes the MSE with an \textit{infinite-capacity fronthaul} \cite{joham2005linear} and is given by $\textbf{P}^{\mathrm{WF}}= \Bar{\alpha}\Bar{\textbf{W}}$, where $\Bar{\textbf{W}}= \textbf{H}^H (\textbf{H}\textbf{H}^H+\frac{KN_0}{q}\textbf{I}_K)^{-1}$ and $\Bar{\alpha} = \sqrt{q / \mathrm{tr} (\Bar{\textbf{W}}\Bar{\textbf{W}}^H) }$. By substituting $\textbf{P}^{\mathrm{WF}}$ into (\ref{eq:betaopt}), we  simplify the expression to \cite{joham2005linear}
\begin{equation}
\beta^{\mathrm{WF}}=\frac{1}{\sqrt{q}} \Big[ \mathrm{tr} \Big( \textbf{H}^H \Big(\textbf{H}\textbf{H}^H+\frac{KN_0}{q}\textbf{I}_K\Big)^{-2}\textbf{H} \Big)\Big]^{1/2}.\label{eq:betawf}
\end{equation}

%%%%%%%%%%%%%%%%%%%%%%%%%%%%%%%%%%%%%%%%%%%
\subsection{Optimization Solution}
\vspace{-2mm}
To find the optimal precoder for an arbitrary fixed $\beta$, first we simplify the objective function of problem (\ref{eq:mmse}) as 
\vspace{-1mm}
\begin{align}
&\mathbb{E}[  \left \Vert \textbf{s} -  \beta \textbf{y}\right \|^2]  = \mathbb{E}[  \left \Vert \textbf{s} -  \beta \textbf{H}\textbf{P}\textbf{s} -  \beta \textbf{n} \right \|^2]  \nonumber
\\  &= \mathrm{tr} \Big( (\textbf{I}_K-\beta \textbf{H}\textbf{P}) \mathbb{E}[\textbf{s}\textbf{s}^H] (\textbf{I}_K-\beta \textbf{H}\textbf{P})^H   +\beta \beta^* \mathbb{E}[ \textbf{n}\textbf{n}^H ] \Big) \nonumber
\\ & = \mathrm{tr} \Big( \textbf{I}_K-\beta \textbf{H}\textbf{P}- \beta^* \textbf{P}^H
 \textbf{H}^H + |\beta|^2\textbf{P}^H\textbf{H}^H\textbf{H}\textbf{P} \Big)  + |\beta|^2 K N_0 \nonumber
 \\ &= \mathrm{tr} \Big(|\beta|^2\textbf{P}^H\textbf{H}^H\textbf{H}\textbf{P} -\beta \textbf{H}\textbf{P}- \beta^* \textbf{P}^H
 \textbf{H}^H  \Big) \!+\! K(|\beta|^2N_0+1). \label{eq:simplified}
\end{align} 
To minimize (\ref{eq:simplified}) with respect to $\textbf{P}$, we can drop the constant term $K(|\beta|^2N_0+1)$ and rewrite the relaxed version of (\ref{eq:mmse}) as 
\begin{mini}|l|
	  {\textbf{P}\in \mathcal{P}^{M \times K}  }{\mathrm{tr}\Big(\textbf{P}^H\textbf{H}^H\textbf{H}\textbf{P}- \frac{1}{\beta^*}\textbf{H}\textbf{P} - (\frac{1}{\beta^*}\textbf{H}\textbf{P})^H \Big)}{}{}
	  \addConstraint{\mathrm{tr}(\textbf{P}\textbf{P}^H) \le q.} \label{eq:prob}
\end{mini}
To turn (\ref{eq:prob}) into a vector optimization problem, we set $\textbf{a} = \mathrm{vec}(\textbf{P}), \textbf{h} = \mathrm{vec}(\frac{1}{\beta^*} \textbf{H}^T)$, so we have
\begin{mini}|l|
	  {\textbf{a}\in \mathcal{P}^{MK \times 1}  }{\textbf{a}^H \left ( \textbf{I}_K \otimes \textbf{H}^H\textbf{H} \right )\textbf{a} -\textbf{h}^T\textbf{a} - \left ( \textbf{h}^T\textbf{a}\right )^H }{}{}
	  \addConstraint{\textbf{a}^H\textbf{a} \le q,} \label{eq:modify}
\end{mini}
where $\otimes$ denotes the Kronecker product. For solving the problem in the quantized domain and reusing  standard MIP results, we should transfer \eqref{eq:modify} into an equivalent real-valued problem utilizing the following definitions:
\begin{align}
 \textbf{a}_\mathbb{R} & =
\begin{bmatrix}
\mathfrak{R}\{\textbf{a}\}\\
\mathfrak{I}\{\textbf{a}\}
\end{bmatrix}, 
\textbf{c}_\mathbb{R}=
\begin{bmatrix}
\mathfrak{R}\{\textbf{h}\}\\
 \mathfrak{I}\{\textbf{h}\} \nonumber
\end{bmatrix}, \text{and } \\
\textbf{V}_\mathbb{R}&=
\begin{bmatrix}
\mathfrak{R}\{\textbf{I}_K \otimes \textbf{H}^H\textbf{H}\} & -\mathfrak{I}\{\textbf{I}_K \otimes \textbf{H}^H\textbf{H}\} \\
\mathfrak{I}\{\textbf{I}_K \otimes \textbf{H}^H\textbf{H}\} & \mathfrak{R}\{\textbf{I}_K \otimes \textbf{H}^H\textbf{H}\}
\end{bmatrix}.
\end{align}
These definitions enable us to rewrite (\ref{eq:modify}) as 
\begin{mini}|l|
	  {\textbf{a}_\mathbb{R}\in \mathcal{\tilde{P}}^{2MK \times 1} }{\textbf{a}^T_\mathbb{R} \textbf{V}_\mathbb{R}\textbf{a}_\mathbb{R} -2\textbf{c}^T_\mathbb{R}\textbf{a}_\mathbb{R} }{}{}
	  \addConstraint{\textbf{a}^T_\mathbb{R}\textbf{a}_\mathbb{R} \le q ,} \label{eq:real}
\end{mini}
where $\mathcal{\tilde{P}}=\mathcal{L}$ assure that we are not using more quantization steps than allowed.
Both the objective function and constraint of (\ref{eq:real}) are convex function of $\textbf{a}_\mathbb{R}$. Due to the $\textbf{a}_\mathbb{R}\in \mathcal{\tilde{P}}^{2MK \times 1}$ criteria, the search domain of the problem is discrete. Numerical algorithms for solving such integer optimization problems to global optimality are well known, e.g., see \cite{wolsey2007mixed}. Hence, we can use standard integer convex solvers to find the optimal solutions efficiently by defining following equivalent problem 
\begin{mini}|l|
	  {\textbf{x} \in \mathbb{Z}^{2MK \times 1} }{\textbf{a}^T_\mathbb{R} \textbf{V}_\mathbb{R}\textbf{a}_\mathbb{R} -2\textbf{c}^T_\mathbb{R}\textbf{a}_\mathbb{R}   }{}{}
	  \addConstraint{\textbf{a}^T_\mathbb{R}\textbf{a}_\mathbb{R} \le q}
	  \addConstraint{\textbf{a}_\mathbb{R}=\Delta \Big( \textbf{x} - (\frac{L-1}{2}) \textbf{1}_{2MK \times 1} \Big)}
	  \addConstraint{\textbf{0}_{2MK \times 1} \le \textbf{x} \le (L-1)\textbf{1}_{2MK \times 1}.} \label{eq:realinteger}
\end{mini}
In the next section, we will use the Gurobi solver with CVX \cite{gurobi2020reference} to solve this problem. Although the complexity of the problem increases exponentially with $MK$, the numerical results show that it is still solvable for practically-sized MU-MIMO systems. The complexity of problem \eqref{eq:realinteger} is also exponential in the number of quantization levels $L$, but we consider a fixed and relatively small number of quantization bits.

%%%%%%%%%%%%%%%%%%%%%%%%%%%%%%%%%%%%%%%%%%%%%%%%%%%%%%%%%%%%%%%%%%%%%%%%%%%%%%%%%
\section{Numerical results}\label{sec5num}

In this section, we compare the sum rates achieved by quantization-aware precoding and quantization-unaware precoding under different conditions. The entries of the channel matrix $\textbf{H}$ are generated as independent circularly-symmetric complex Gaussian random variables with variance $\gamma$ and the common SNR of all UEs is defined as $\mathrm{SNR}=\frac{q \gamma}{N_0}$. The number of BS antennas is $M=16$ and the number of UEs is $K=4$. The  sum rate is calculated using Monte Carlo simulations for the case of Gaussian signaling and perfect CSI at the receiver. We will compare different precoding schemes as a function of the SNR and the number of quantization levels $L$. We will compare quantization-based precoding with the infinite-resolution case. The classic WF precoding and maximum ratio transmission (MRT) schemes are considered.

\begin{figure}[!t]
        \centering
        \includegraphics[scale=0.45]{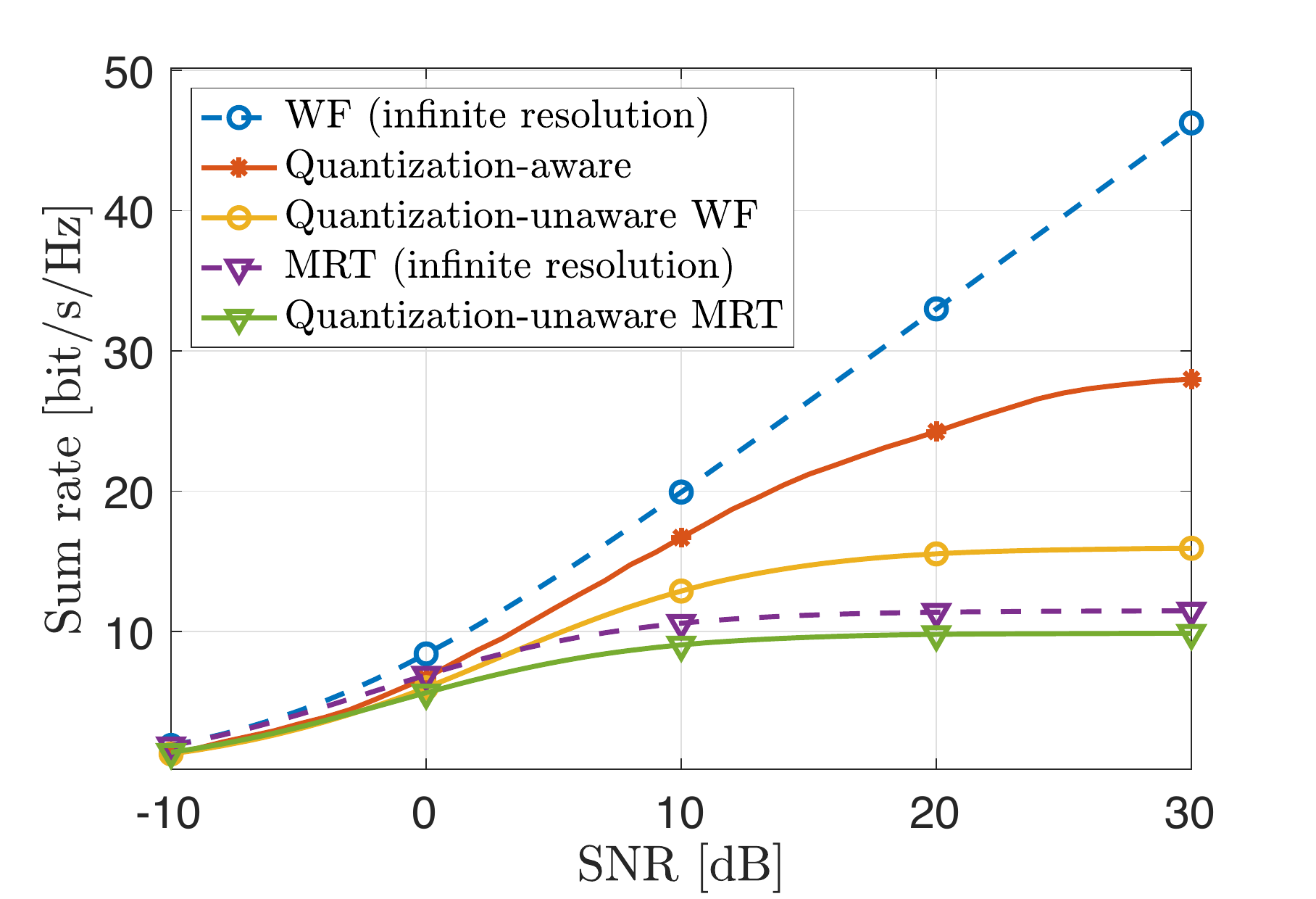}
        \caption{Average achievable sum rate versus the SNR for different precoding schemes.}
        \label{fig:sum}
\end{figure} 

Fig.~\ref{fig:sum} depicts the average sum rate as a function of the SNR for different precoding schemes: quantization-aware, quantization-unaware, and infinite resolution. The number of quantization levels is $L=8$. The infinite-resolution WF precoding outperforms all the quantized precoding schemes and the gap increases linearly (in dB scale) at high SNR. However, the gap between the quantization-aware and infinite-resolution WF precoding is  remarkably smaller than the gap between quantization-unaware WF and the infinite-resolution WF precoding. The quantization-aware and -unware MRT precoding have the same performance as WF at low SNR, but the lack of interference cancellation results in a large gap at higher SNRs. The gap between the MRT curves corresponds to the loss in beamforming gain due to quantization. We notice that for quantization-aware and quantization-unaware precoding, the sum rate converges to specific limits at high SNR, since all interference cannot be canceled due to the quantization effect; that is, the system is interference-limited at high SNR. 

\begin{figure}[!t]
        \centering
        \includegraphics[width=\columnwidth]{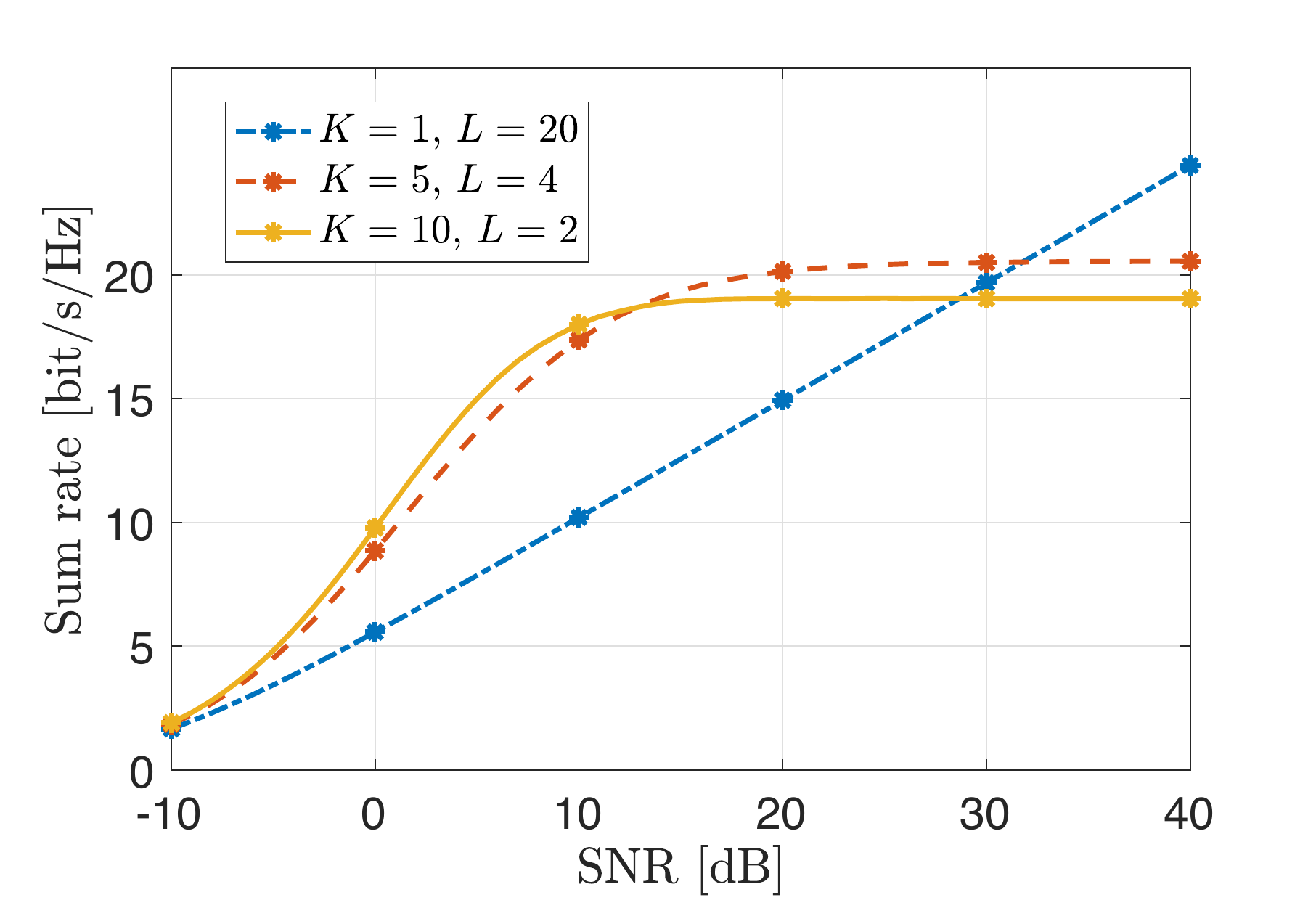}
        \caption{Average achievable sum rate versus the SNR for a fixed number of BS antennas.}
        \label{fig:fixed}
\end{figure}

Similar to \cite{jindal2006mimo}, the degrees-of-freedom of the considered MU-MIMO system is $1$, due to the finite-resolution quantization. This implies that it is optimal to serve one UE at a time when the SNR is large. To demonstrate this, Fig.~\ref{fig:fixed} presents the average achievable sum rate as a function of SNR for $M=16$ BS antennas and different values of $K$ and $L$, such that $K \cdot L = 20$. At low and medium SNR, the sum rate is maximized by serving many UEs. At high SNR, serving a lower number of UEs with a high-resolution quantizer outperforms the opposite case. This is because the system is heavily interference-limited, which can be partially resolved by increasing $L$.
For each SNR value, there is an optimal number of UEs to serve and it is imperative to schedule the right number of UEs.
Despite the fact that single-user transmission prevails at high SNR, MU-MIMO remains the preferable case in practice since the crossing point is at 20 bit/s/Hz, which would require enormous constellations for a single UE.

%%%%%%%%%%%%%%%%%%%%%%%%%%%%%%%%%%%%%%%%%%
%%%%%%%%%%%%%%%%%%%%%%%%%%%%%%%%%
\section{Conclusions and future work} \label{sec6conc}

5G sites consist of an AAS connected to a BBU via a digital fronthaul with limited capacity. In the downlink, the finite-constellation data symbols can be sent to the AAS without quantization, but the precoder matrix must be quantized to finite precision. We have introduced the concept of novel quantization-aware precoding, where the BBU uses the quantizer structure to select the best finite-precision MU-MIMO precoder that requires no further quantization. In particular, we formulate the MSE-minimizing precoder and solve it by MIP.
We have shown numerically that the proposed quantization-aware precoding outperforms the baseline 
quantization-unaware precoding, where the optimal precoding for the infinite-resolution case is selected and then quantized. The improved interference mitigation gives a large sum rate gain at medium and large SNRs, despite the fact that the degrees-of-freedom is limited to one.

We will develop lower-complexity quantization-aware precoders in future work to make the concept useful also in massive MU-MIMO scenarios.

% -------------------------------------------------------------------------
\bibliographystyle{IEEEbib} \bibliography{IEEEabrv,references}

\end{document}